# A Perceptual Aesthetics Measure for 3D Shapes

Kapil Dev[1]   Manfred Lau[1]   Ligang Liu[2]
[1]Lancaster University    [2]USTC China

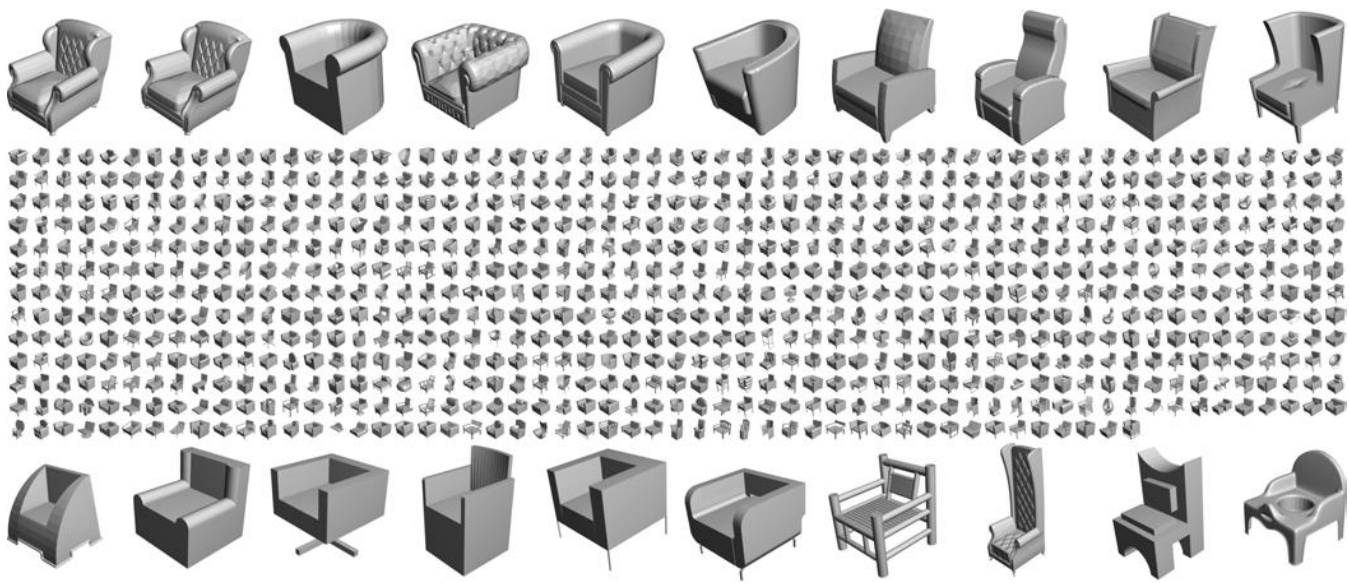

**Figure 1:** *778 "club chair" 3D models downloaded online and ranked by our aesthetics measure (please zoom in to see shape details). The rankings are from the top to bottom rows and left to right in each row. The top and bottom 10 models are enlarged for easier visualization.*

## Abstract

While the problem of image aesthetics has been well explored, the study of 3D shape aesthetics has focused on specific manually defined features. In this paper, we learn an aesthetics measure for 3D shapes autonomously from raw voxel data and without manually-crafted features by leveraging the strength of deep learning. We collect data from humans on their aesthetics preferences for various 3D shape classes. We take a deep convolutional 3D shape ranking approach to compute a measure that gives an aesthetics score for a 3D shape. We demonstrate our approach with various types of shapes and for applications such as aesthetics-based visualization, search, and scene composition.

## 1 Introduction

The word "aesthetics" generally refers to the nature of beauty and has traditionally been considered in art, music, and poetry. In these fields, it has been postulated that mathematical criterias such as minimum description length and self-similarity [Peitgen and Richter 1986] tend to evoke aesthetics preferences. In this paper, we focus on the aesthetics of 3D shapes, where we consider aesthetic shapes as those that are visually attractive or pleasing. Many 3D shapes around us trigger a universal pleasing response irrespective of our backgrounds and experiences [Séquin 2005; Fiore 2010; Gambino 2013]. For a consumer, the form of a product plays an important role in the decision towards purchasing the product. A computational understanding of shape aesthetics allows us to more effectively use large datasets of 3D shapes.

There has been much previous work that explores the aesthetics of images [Liu et al. 2010; Redies et al. 2012]. In particular, the attractiveness and beautification of human faces in images have been explored [Eisenthal et al. 2006; Leyvand et al. 2008; Said and Todorov 2011]. For 3D shapes, the study of aesthetics [Séquin 2005; Bergen and Ross 2012; Vartanian et al. 2013; Miura and Gobithaasan 2014] has considered specific manually-crafted features such as curvature and symmetry and mathematical properties such as bending energy and minimum variation surface. However, these are pre-defined features in the sense that curvature, for example, is pre-selected as a feature to be tested. In contrast, our approach works *without* manually-crafted features such that there is no bias or pre-defined conception of how to computationally define aesthetics. We leverage one of the fundamental concepts of deep learning to do so. We learn directly from raw voxel data and autonomously learn an aesthetics measure to best fit with human aesthetics preferences.

Comparing the aesthetics of shapes can be difficult if the shapes to be compared against are quite different. As a metaphor, we try to avoid comparing between "apples and oranges" but wish to compare apples with apples. We use 3D models from the ShapeNet dataset [Chang et al. 2015] which are already classified into human-understandable categories of man-made objects such as club chairs, mugs, and lamps. Hence we use these categories and only compare models from within each category.

We consider aesthetics as a perceptual concept and we collect a large amount data on the shape aesthetics preferences from humans and learn from the data. We strive to take a simple but effective data collection method. It is relatively difficult for humans to give an absolute aesthetics score to a single shape. Inspired by [Garces et al. 2014; Liu et al. 2015; Lun et al. 2015; Lau et al. 2016], we instead ask humans to compare between two shapes and select the one that they believe to be more aesthetic. The shapes are also rotated to show views from different directions. We require participants to pick one of the shapes from each pair. From our experiences, this requirement encourages them to think about the shapes and provide good responses, rather than allowing them the option to not provide a response in some cases. Our work will show that these pairs of shapes are enough for learning shape aesthetics. As crowdsourcing has been a common approach for data collection for various graphics problems [Gingold et al. 2012b; Gingold et al. 2012a; Garces et al. 2014], we also use a crowdsourcing framework and post the pairs of 3D shapes as tasks on Amazon Mechanical Turk. The collected data about aesthetics preferences may depend on each individual's personal culture and background. We analyze the collected data by checking for consistency at the platform-level, population-level, and individual-level.

We take a deep convolutional 3D shape ranking approach to compute our aesthetics measure. The motivation for a learning-to-rank based approach is that our collected data is ranking-based (e.g. one shape is more aesthetic than another). The deep architecture allows us to autonomously learn features from raw voxel data. We take 3D meshes of each object type as input and voxelize them. The voxel representation works well for our problem as shape features can be autonomously computed from the representation. The raw voxel data is the first layer to a deep neural network that computes an aesthetics score for each mesh. To learn the weights for the network with our ranking-based data, we use a deep convolutional ranking formulation and backpropagation that uses two copies of the deep neural network. After training the network, we can use one copy of the learned network to compute an aesthetics score for a new 3D mesh of the corresponding object type.

We demonstrate our aesthetics measure by collecting a large number of models from the ShapeNet dataset [Chang et al. 2015] and ranking the models in each category based on their aesthetics scores. We consider only the shape geometry which is already interesting and not the color or texture. We analyze the prediction accuracy of the neural network and compare the learned measure with specific aesthetics criterias considered in previous work. Finally, we show how our aesthetics measure can be applied to produce better tools for visualization, 3D model search, and scene composition.

The contributions of this paper are: (1) We explore the problem of 3D shape aesthetics by learning with raw voxel data and *without* any pre-defined conception of a computational description of aesthetics; (2) We solve this problem with a deep convolutional 3D shape ranking approach; and (3) We evaluate our 3D shape aesthetics measure and demonstrate the applications of aesthetics-based visualization, search, and scene composition.

## 2 Related Work

### 2.1 Shape Features for Aesthetics

**Shape Features.** From a computational perspective, previous work has mathematically defined functions and geometric features for shape aesthetics. Séquin [2005] introduces the idea of "optimizing a surface by maximizing some beauty functional." His work focuses on abstract sculptural forms for artistic purposes and he mathematically defines "beauty functionals" that have properties that lead to more beautiful shapes. A study of the link between a user's emotional reactions and a product's basic geometric elements has been performed [Giannini and Monti 2002]. Their work has been applied to the automotive and household supplies fields. A fuzzy shape specification system uses pre-defined aesthetic descriptors for designing shapes [Pham and Zhang 2003]. Symmetry [Mitra et al. 2007] is a feature that has been associated with aesthetic shapes [Bergen and Ross 2012]. Specific mathematical criterias including entropy, complexity, deviation from normality, 1/f noise, and symmetry have been tested for creating aesthetic 3D models for evolutionary art [Bergen and Ross 2012].

From a product design perspective, elements of design including color, light, line and shape, texture, and space and movement are considered useful for understanding aesthetics by designers [Fiore 2010]. Specific aesthetic features have also been considered for modeling shapes for product design [Fujita et al. 1999].

**Curvature as an Appealing Shape Feature.** There is one type of shape feature that has been particularly identified as being important towards shape aesthetics. This feature is curvature: in general, a shape that has more curved parts or surfaces tends to be more appealing. For example, a study that uses functional magnetic resonance imaging asks participants to judge whether architectural spaces are beautiful [Vartanian et al. 2013]. Participants were more likely to judge them as beautiful if they were curvilinear than rectilinear. Another study asks art gallery visitors to observe an image set containing shape variations of a sculpture [Gambino 2013]. Participants were asked to note their "most preferred" and "least preferred" shapes on a ballot. This experiment found that the visitors prefer shapes with gentle curves as opposed to those with sharp points. Specific shape criterias such as fairness metric, bending energy, and minimum variation surface have been considered as a way to describe the relation between curvature and aesthetic surfaces [Miura and Gobithaasan 2014].

**Shape Aesthetics in Specific Applications.** There is previous work in exploring specific object types and applications, and defining specific shape features for them. For building exteriors and urban design, high-level features that determine the aesthetic quality of buildings and their surroundings have been studied [Nasar 1994]. For office chair design, researchers have considered user satisfaction criterias such as luxuriousness, balance, and attractiveness [Park and Han 2004]. They build a fuzzy rule-based model based on specific variables that are related to these user satisfaction criterias. For jewelry design, the aesthetics of jewelry shapes have been considered [Wannarumon 2010]. They allow the user to adjust specific shape features such as golden ratio, mirror symmetry, and rotational symmetry to design more aesthetic shapes.

Compared to previous work related to shape features for aesthetics, the key difference in our work is that we learn shape features automatically from raw voxel data without pre-defining any aesthetic features or mathematical functions. We show that our learned aesthetics measure can separate between shapes with curved and planar surfaces, although there is no mathematical description of curvature pre-defined in our method and we were not looking for any shape features beforehand. Our method can work with general shapes as long as we have the data for each class of shapes.

### 2.2 Perception of Shape Aesthetics

Shape aesthetics based on human perception and human factors have been considered in different ways. There is work that explores product design from a human factor perspective [Jordan 2002]. The book discusses human pleasures and consumer needs for product design. Howlett et al. [2005] study the visual fidelity of polygonal models. Secord et al. [2011] find good views of a 3D object by optimizing the parameters of a perceptual model for viewpoint goodness. The perceptual model is optimized by data collected from a large user study. Zhang et al. [2015] develop a perceptual model for 3D printing direction. Their method finds printing directions that avoid placing printed supports in perceptually salient areas, thereby reducing the visual artifacts caused by the printed supports. Chew et al. [2016] directly measure the aesthetic perception of virtual 3D shapes with electroencephalogram (EEG) signals. Our work is different in developing a generic aesthetics measure for 3D shapes. We solve this problem with a large-scale aesthetics data collection process and a deep ranking based approach.

### 2.3 Shape Aesthetics With User Input

The idea of generating many variations of a shape for the user to choose from has been explored in previous work. An existing system [Smyth and Wallace 2000] has the user start with a shape or form, and the system generates variations of it. The user then selects the appealing ones and the process is repeated. Another system [Lewis and Ruston 2005] takes an evolutionary framework. The user chooses the most appealing designs from a set of options and the system generates a new population of designs by combining and modifying the user-selected shapes. This process is repeated to generate aesthetic geometric designs. These previous works let the users iteratively choose their preferred shapes and our key difference is that we take a learning framework.

### 2.4 Image Aesthetics and Human Face Attractiveness

**Image Aesthetics.** There is much previous work in the aesthetics of images. For example, existing works have explored the aesthetic quality of pictures based on their visual content [Datta et al. 2006], the optimal positions of objects in images [Liu et al. 2010], and the attractiveness of images and paintings [Redies et al. 2012].

**Facial Images and Shapes.** In particular, attractiveness has been well studied for facial images and shapes. Eisenthal et al. [2006] train a predictor for the attractiveness of face images based on the image pixels and the specific proportions of facial features. Leyvand et al. [2008] take forward-facing face images and beautify them based on editing various facial feature locations. Said et al. [2011] identify components of attractiveness in face images. For 3D face shapes, O'Toole et al. [1999] find that an average face shape and average texture leads to a more attractive face shape. Liao et al. [2012] enhance a 3D face model by performing symmetrization and adjusting various facial proportions based on the golden ratio. While previous work exists for 2D images and photos, and for 3D face shapes where specific face features such as averageness, symmetry, and golden ratio are used, the difference in our work is in an aesthetics measure for general 3D shapes and without any pre-defined features.

### 2.5 Crowdsourcing

Crowdsourcing has recently been used for solving various graphics problems. Chen et al. [2012] ask users to select points on the surface of a mesh that they think will also be selected by others. These "Schelling points" are points of interest that can be used for mesh simplification. Garces et al. [2014] learn a similarity measure of style for 2D clip art with crowdsourced data. Liu et al. [2015] learn a measure of style compatability for furniture models. Lun et al. [2015] use crowdsourcing data to study the perceptual style similarity of 3D shapes. These previous works compute various 2D image features or 3D shape descriptors and learn with these features. The key difference in our work is that we do not pre-specify features but learn them autonomously.

### 2.6 Deep Learning

**Deep Ranking.** Previous works have used the concept of deep ranking for computer vision problems. For example, a deep ranking approach is used to compute image similarity [Wang et al. 2014]. A deep metric learning approach is used to compare between face images [Hu et al. 2014]. A convolutional network architecture can be applied to comparing between image patches [Zagoruyko and Komodakis 2015].

**Deep Learning for 3D Modeling.** Recently, deep learning has been applied for solving 3D modeling problems. These include computing human body correspondences [Wei et al. 2015], 3D shape recognition [Su et al. 2015], and tactile mesh saliency [Lau et al. 2016]. We also apply the concept of deep ranking in our work

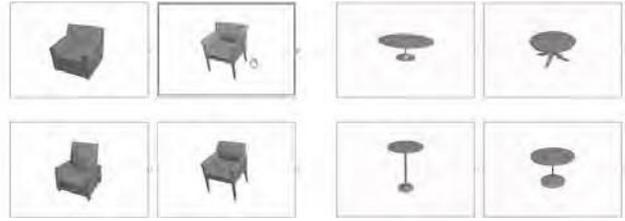

**Figure 2:** *Examples of HIT tasks or queries (for club chairs and pedestal tables) posted on Amazon Mechanical Turk. In each of the four tasks, the user selects one shape as being more aesthetic than the other. Each 3D shape is shown in an image where the shape is iteratively rotated (to provide 3D views) and paused for a short time (in a front-facing view that we chose that best displays each class of shapes).*

and we use it to solve a different problem of computing a perceptual 3D shape aesthetics measure.

## 3 Collecting Aesthetics Data

Our overall framework is to collect data on the human perception of visual shape aesthetics and then learn from the data to get an aesthetics measure. This section describes the data collection process and the testing for data consistency. Our work is inspired by the collection of triplets data that describes the style similarity of 2D clip art [Garces et al. 2014] and 3D shapes [Liu et al. 2015; Lun et al. 2015]. In our work, the key data collection step is to collect data regarding pairs of 3D shapes for each class separately.

We collect 3D models from the ShapeNet dataset [Chang et al. 2015]. In the dataset, the models are labeled into human-understandable categories. Each model is already rotated and scaled correspondingly with the other models in the same category. Table 1 provides some information about the shapes we used.

| Class of 3D Shapes | Num | $|I_{train}|$ | $|I_{validation}|$ |
|---|---|---|---|
| Club Chairs | 778 | 7600 | 400 |
| Pedestal Tables | 40 | 2578 | 297 |
| Mugs | 75 | 743 | 82 |
| Lamps | 88 | 2250 | 250 |
| Dining Chairs | 277 | 4790 | 310 |

**Table 1:** *"Num" is the total number of shapes in each class. We separate the total number of samples in each class into a training data set $I_{train}$ and a validation data set $I_{validation}$. $|I|$ is the number of samples in $I$.*

Each data sample consists of a pair of shapes of the same class and a human selection of the more aesthetic shape in the pair. We have 30 such samples or tasks in each HIT (i.e. a set of tasks posted on Amazon Mechanical Turk). Figure 2 shows some example HIT tasks. The participants or Turkers are asked to select one shape from each pair that is more attractive to them. They must select one shape over the other and there is no choice to indicate that they may be similar in aesthetics. We believe that this allows the participants to actively think about the shapes, rather than going through the tasks and just randomly clicking their responses. Participants were paid $0.10 for each HIT.

We realize that for some pairs of shapes, there may be no "right or wrong" answers and it depends on the user preferences. Hence the crowdsourced data may not be reliable. We use various methods *during the data collection* to attempt to get good data: (i) We provide clear instructions to tell the Turkers that unhonest workers (based on information provided on Amazon Mechanical Turk) will be blocked; (ii) After users click on a response, we have a 4 second delay before they can click on the next response; and (iii) We include control questions as in previous work [Garces et al. 2014] where one shape from the pair is intentionally made to be ugly (see Figure 6 for some examples). For each HIT, we have five control questions and the user must correctly answer all of them for us to accept the tasks in the HIT.

|              |   | P1   | P2   | P3   | P4   | P5   | P6    | P7   | P8   | P9   | P10  |
|--------------|---|------|------|------|------|------|-------|------|------|------|------|
| Volunteer    | A | 20.0 | 33.3 | 26.7 | 53.3 | 60.0 | 26.7  | 40.0 | 33.3 | 53.3 | 46.7 |
|              | B | 80.0 | 66.7 | 73.3 | 46.7 | 40.0 | 73.3  | 60.0 | 66.7 | 46.7 | 53.3 |
| Paid         | A | 46.7 | 73.3 | 26.7 | 73.3 | 53.3 | 33.3  | 13.3 | 33.3 | 40.0 | 46.7 |
|              | B | 53.3 | 26.7 | 73.3 | 26.7 | 46.7 | 66.7  | 86.7 | 66.7 | 60.0 | 53.3 |

|        |   | P1   | P2   | P3   | P4   | P5   | P6    | P7   | P8   | P9   | P10  |
|--------|---|------|------|------|------|------|-------|------|------|------|------|
| Male   | A | 70.0 | 80.0 | 70.0 | 50.0 | 40.0 | 0.0   | 60.0 | 35.0 | 75.0 | 80.0 |
|        | B | 30.0 | 20.0 | 30.0 | 50.0 | 60.0 | 100.0 | 40.0 | 65.0 | 25.0 | 20.0 |
| Female | A | 68.2 | 59.1 | 68.2 | 45.5 | 59.1 | 0.0   | 77.3 | 50.0 | 81.8 | 86.4 |
|        | B | 31.8 | 40.9 | 31.8 | 54.5 | 40.9 | 100.0 | 22.7 | 50.0 | 18.2 | 13.6 |

|                |   | P1   | P2   | P3   | P4   | P5   | P6   | P7   | P8   | P9   | P10  |
|----------------|---|------|------|------|------|------|------|------|------|------|------|
| Random Group 1 | A | 9.5  | 14.3 | 23.8 | 47.6 | 66.7 | 19.0 | 57.1 | 23.8 | 19.0 | 28.6 |
|                | B | 90.5 | 85.7 | 76.2 | 52.4 | 33.3 | 81.0 | 42.9 | 76.2 | 81.0 | 71.4 |
| Random Group 2 | A | 47.6 | 42.9 | 33.3 | 52.4 | 33.3 | 38.1 | 52.4 | 14.3 | 19.0 | 28.6 |
|                | B | 52.4 | 57.1 | 66.7 | 47.6 | 66.7 | 61.9 | 47.6 | 85.7 | 81.0 | 71.4 |

**Figure 3:** *Testing for Data Consistency. The 3 tables show example results for platform-level (volunteer or paid participants), population-level (male or female), and individual-level (participants randomly split into 2 groups) consistency respectively. (P1, P2, ...) are pairs of shapes (A,B) from which the participants can choose A or B as being more aesthetic. The numbers shown are the percentages of participants who selected each response.*

At the start of each HIT, we also collect some demographics data from the participants. This data includes their gender, age group, and region. We had 403 male and 360 female Turkers (and 12 who did not provide their gender). The HIT acceptance rates based on gender are 87.1% for males and 82.8% for females. We had the following age groups: (0-20, 21-30, 31-40, 41-50, 51-60, 60-100) and the percentages of Turkers in each group respectively are: (1.6%, 36.0%, 37.1%, 14.3%, 9.6%, 1.3%). We had the following regions: (Africa, Asia, Australia, Europe, North America, South America). The HIT acceptance rates based on region are: (N/A due to no Turkers, 85.1%, 100%, 87.9%, 85.0%, 77.3%).

### 3.1 Testing for Data Consistency

*After the data collection* and before using the data for training, we wish to analyze the robustness of the data. The motivation is that it is difficult to know the accuracy of Turker responses as there is no ground truth to compare against. Hence we essentially compare the data against itself. We realize that shape aesthetics may depend on personal preferences and backgrounds. In addition, even the same shape pair given to different participants can lead to different but "correct" responses. We thereby analyze the data for three types of consistency.

**Platform-Level Consistency.** The collected data may depend on whether participants are volunteers or paid. For example, Redi and Povao [2014] find differences between volunteer and paid participants in crowdsourcing tasks. We compare between posting tasks on the platform of Facebook (volunteer) and Amazon Mechanical Turk (paid). We have 25 shape pairs x 15 participants who provided responses on each platform (where each platform has a different set of participants). The first table in Figure 3 shows some example results (i.e 10 shape pairs). We consider the responses for each shape pair to match if the "volunteer majority (i.e. greater than or equal to 50%) response" is the same as the "paid majority response." The results show 21 out of 25 matches. This is evidence that the data is consistent across the platforms. We choose to use a paid crowdsourcing platform as it is more convenient for collecting large amounts of data.

**Population-Level Consistency.** A participant's preferences on shape aesthetics may depend upon his/her cultural background and personal experiences. For our tasks, users can enter demographics information and we analyze the data based on them. We collected data for 25 shape pairs x 50 participants. 6 users were rejected and 2 users did not provide demographics data, leaving a total of 42 users. We split the data into responses provided by male or female users. The second table in Figure 3 shows some example results. The results show 21 of 25 matches. We also split the data into responses from Asia and North America. Other regions such as Africa and Australia had no or too few users. For this case, we also have 21 of 25 matches. This provides evidence that the data is consistent at the population level.

**Individual-Level Consistency.** Shape aesthetics may also be dependent on individual preferences. One participant can choose A while the other chooses B and both may have their reasons for doing so. We use the same data collected for the population-level consistency tests. We evenly split the 42 users into two random groups. The third table in Figure 3 shows some example results. We have 21 out of 25 matches for this case. We then perform another random split of the data into two groups of users, and we get 22 out of 25 matches. Hence there is evidence that the data is consistent across individuals.

As the data is consistent on all three levels, we do not split the collected data according to demographics for the learning process.

## 4 Deep Convolutional 3D Shape Ranking

This section describes how we learn an aesthetics measure from the collected data described in the previous section. We take a deep convolutional 3D shape ranking approach. The motivation for each of these terms is as follows. We take a *deep* multi-layer neural network architecture to learn a potentially complex and non-linear relationship from the data to the aesthetics scores. The *convolutional* architecture is needed when the voxel resolution is high. We use voxels to represent *3D shapes* which has been shown to be an effective representation for deep learning [Wu et al. 2015] and we experimented with voxels at different resolutions as input to the neural network. As our collected data is based on *rankings* of pairs of 3D shapes, we use a learning technique commonly known as learning-to-rank [Parikh and Grauman 2011] to compute an overall measure that "best fits" with the paired rankings in the data.

As we collect pairs of data, we take a deep ranking formulation that is inspired by [Hu et al. 2014; Zagoruyko and Komodakis 2015; Lau et al. 2016] and that fits well with our collected data and problem. The main differences of our method include: our data compares between pairs of 3D shapes instead of points on the shapes, and we convert the shapes into their voxel representations instead of working with images taken from different viewpoints of the shapes. If the voxel resolution is high, we use a convolutional neural network architecture as otherwise a fully-connected network would not be practical to train. We experimented with various voxel resolutions and neural network architectures to gain insight into what works well for this shape aesthetics problem. Furthermore, we have two copies of the neural network (which takes the concept of Siamese networks [Bell and Bala 2015]) instead of four copies in the backpropagation.

We first describe the voxel data representation and the neural network architectures. We next describe the deep ranking formulation and the backpropagation in the neural network that works with the collected data pairs. After the training process, we have an aesthetics measure that gives a score for each 3D shape.

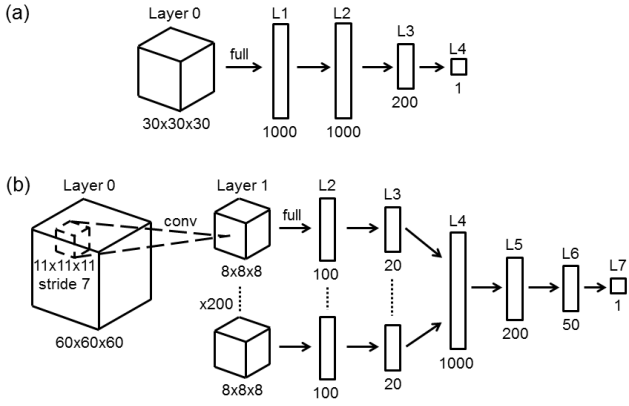

**Figure 4:** *Our deep neural network architectures. The input in layer 0 is the voxel representation of a 3D shape and the output in the last layer is the shape's aesthetics score. In each case, we need two copies of the network to compute the partial derivatives for the batch gradient descent. (a) For low resolution voxels such as 15 and 30, we can have a fully-connected network. The number of nodes is indicated for each layer. (b) At higher resolutions such as 45 and 60, it is more practical to use a convolutional architecture. There is one convolutional layer, and then from layer 1 onwards we use a fully-connected network as the number of nodes becomes relatively small.*

## 4.1 Voxel Data Representation and Deep Neural Network Architectures

We choose a voxel representation as this is a basic representation from which more complex features may be computed. We voxelize each mesh and the voxels become the input to the first layer of the deep neural network.

We experimented with different neural network architectures (Figure 4). We can have a low resolution voxel representation, where the nodes between each successive layers are fully-connected. As we increase the voxel resolution, we need a convolutional architecture. We do not use any pooling layers as we wish to keep the details of the shapes in the voxel representation. The motivation for specifically experimenting with different resolutions is that when humans make decisions on shape aesthetics, we may observe the overall shape and this corresponds to a lower resolution and fully-connected layers for the whole shape (Figure 4a). We may also observe the details of the shape and this corresponds to a higher resolution and some convolutional layers to recognize more local features of the shape (Figure 4b).

We let $\mathbf{W}$ be the set of all weights consisting of $\mathbf{W}^{(l)}$ between each successive layers, where $\mathbf{W}^{(l)}$ is the matrix of weights for the connections between layers $l-1$ and $l$. We use the neural network in Figure 4b to provide some examples. In this case, $\mathbf{W}^{(6)}$ has 50x200 values. Between layers 2 and 3, there are multiple (e.g. 200 in this case) sets of such weights. For the convolutional layer between layers 0 and 1, there are 1331x200 weight values. We let $\mathbf{b}$ be the set of all biases consisting of $\mathbf{b}^{(l)}$ for each layer except for layer 0, where $\mathbf{b}^{(l)}$ is the vector of biases for the connections to layer $l$. For example, $\mathbf{b}^{(6)}$ has 50x1 values. Between layers 2 and 3, there are multiple sets of such biases. For the convolutional layer between layers 0 and 1, there are 1x200 bias values.

## 4.2 Deep Ranking Formulation and Backpropagation

Our algorithm takes the set $I_{train}$ and learns a deep neural network that maps the voxels of a 3D shape $\mathbf{x}$ to the shape's aesthetics score $y = h_{\mathbf{w},\mathbf{b}}(\mathbf{x})$ (Figure 4). Since we do not wish to compare "apples with oranges" in our metaphor, each class of shapes is learned separately and has a different network. We follow the deep ranking formulation in [Lau et al. 2016], but there are many subtle and important differences including: the 3D shape and voxel representation, the formulation for the 3D convolutional architecture, and the two copies of the neural network for the $A$ and $B$ cases.

While supervised learning frameworks have the target values $y$ in the training data, we do not directly have such target values. Our data is ranking-based and provides rankings of pairs of 3D shapes. This is the motivation for taking a learning-to-rank formulation and we learn $\mathbf{W}$ and $\mathbf{b}$ to minimize this ranking loss function:

$$\mathcal{L}(\mathbf{W}, \mathbf{b}) = \frac{1}{2}\|\mathbf{W}\|_2^2 + \frac{C_p}{|\mathcal{I}_{train}|} \sum_{(\mathbf{x}_A, \mathbf{x}_B) \in \mathcal{I}_{train}} l(y_A - y_B) \quad (1)$$

where $\|\mathbf{W}\|_2^2$ is the $L^2$ regularizer (2-norm for matrix) to prevent over-fitting, $C_p$ is a parameter, $|I_{train}|$ is the number of elements in $I_{train}$, $l(t) = \max(0, 1-t)^2$ is a suitable loss function for the inequality constraints, and $y_A = h_{\mathbf{w},\mathbf{b}}(\mathbf{x}_A)$.

The training set $I_{train}$ contains inequality constraints. If $(\mathbf{x}_A, \mathbf{x}_B) \in I_{train}$, our neural network should give a higher aesthetics score for shape $A$ than for shape $B$ (i.e. $h(\mathbf{x}_A)$ should be greater than $h(\mathbf{x}_B)$). The loss function $l(t)$ enforces prescribed inequalities in $I_{train}$ with a standard margin of 1.

To minimize $L(\mathbf{W}, \mathbf{b})$, we perform an end-to-end neural network backpropagation with batch gradient descent. First, we have a forward propagation step that takes each pair $(\mathbf{x}_A, \mathbf{x}_B) \in I_{train}$ and propagates $\mathbf{x}_A$ and $\mathbf{x}_B$ through the network with the current $(\mathbf{W}, \mathbf{b})$ to get $y_A$ and $y_B$ respectively. Hence there are two copies of the network for each of the $A$ and $B$ cases. Note that in some cases there are multiple sets of weights and biases between layers and the forward propagation proceeds as usual between each set of corresponding nodes. In the convolutional layer, the same weights and biases in each 3D convolutional mask are forward propagated multiple times.

We then perform a backward propagation step for each of the two copies of the network and compute these delta ($\delta$) values:

$$\delta_i^{(n_l)} = 1 - y^2 \quad \text{for output layer} \quad (2)$$

$$\delta_i^{(l)} = \left(\sum_{k=1}^{s_{l+1}} \delta_k^{(l+1)} w_{ki}^{(l+1)}\right)(1 - (a_i^{(l)})^2) \quad \text{for inner layers} \quad (3)$$

where the $\delta$ and $y$ values are indexed as $\delta_{Ai}$ and $y_A$ in the case for $A$. The index $i$ in $\delta$ is the neuron in the corresponding layer and there is only one node in our output layers. $n_l$ is the number of layers, $s_{l+1}$ is the number of neurons in layer $l+1$, $w_{ki}^{(l+1)}$ is the weight for the connection between neuron $i$ in layer $l$ and neuron $k$ in layer $(l+1)$, and $a_i^{(l)}$ is the output after the activation function for neuron $i$ in layer $l$. We use the *tanh* activation function which leads to these $\delta$ formulas. Because of the learning-to-rank aspect, we define these $\delta$ to be different from the usual $\delta$ in the standard neural network backpropagation. Note that in some cases there are multiple sets of weights and biases between layers and the backward propagation proceeds as usual between each set of corresponding nodes. The backward propagation computes these $\delta$ values from the last layer up to layer 1.

We now compute the partial derivatives for the gradient descent. For $\frac{\partial \mathcal{L}}{\partial w_{ij}^{(l)}}$, we split this into a $\frac{\partial \mathcal{L}}{\partial \|\mathbf{W}\|_2} \frac{\partial \|\mathbf{W}\|_2}{\partial w_{ij}^{(l)}}$ term and $\frac{\partial \mathcal{L}}{\partial y} \frac{\partial y}{\partial w_{ij}^{(l)}}$ terms (a term for each $y_A$ and each $y_B$ computed from each $(\mathbf{x}_A, \mathbf{x}_B)$ pair). The $\frac{\partial \mathcal{L}}{\partial y} \frac{\partial y}{\partial w_{ij}^{(l)}}$ term is expanded for the $A$ case for example to $\frac{\partial \mathcal{L}}{\partial y_A} \frac{\partial y_A}{\partial a_1} \frac{\partial a_1}{\partial z_1} \frac{\partial z_1}{\partial w_{ij}^{(l)}}$ where the last three partial derivatives are computed with the copy of the network for the *A* case. $z_i$ is the value of a neuron before the activation function. The partial derivatives are then computed similar to those in [Lau et al. 2016].

The batch gradient descent starts by initializing **W** and **b** randomly. We then go through the training data for a fixed number of iterations, where each iteration involves taking a set of data pairs and performing the forward and backward propagation steps and computing the partial derivatives. Each iteration of batch gradient descent sums the partial derivatives from a set of data pairs and updates **W** and **b** with a learning rate $\alpha$: using $w_{ij}^{(l)} = w_{ij}^{(l)} - \alpha \frac{\partial \mathcal{L}}{\partial w_{ij}^{(l)}}$ and $b_i^{(l)} = b_i^{(l)} - \alpha \frac{\partial \mathcal{L}}{\partial b_i^{(l)}}$.

### 4.3 Learned Aesthetics Measure

After the batch gradient descent learns **W** and **b**, we can use them to compute an aesthetics score for a 3D shape. For a new shape of the corresponding category, we voxelize it into **x** and use one copy of the neural network and a forward propagation pass to get the score $h_{\mathbf{W},\mathbf{b}}(\mathbf{x})$. This score is an absolute value, but since the data and method are ranking-based, it has more meaning in a relative sense when the score of a shape is compared to that of another.

## 5 Results

We demonstrate our learned aesthetics measure by showing the rankings of a large number of various classes of 3D shapes based on their aesthetics scores. Our aesthetics measure is learned from crowdsourced data and contains the collective preferences of many people. A single score for one shape is not meaningful, while the scores for multiple shapes can be compared against each other to give information about their relative aesthetics.

### 5.1 Validation Data Sets

We use a validation dataset to set the parameters of the neural network. For each category, we keep about 5 to 10% of the collected data as a separate validation set $I_{validation}$ which has the same format as the training data $I_{train}$. For each pair of shapes in $I_{validation}$, the prediction from the measure learned with $I_{train}$ is correct if the collected data says shape A is more aesthetic than shape B and our score of shape A is greater than that of B. To select the parameter for $\alpha$, for example, we can let $\alpha$ be $\{1e^{-1}, 1e^{-2}, 1e^{-3}, 1e^{-4}, 1e^{-5}\}$. The selected $\alpha$ is the one that minimizes the validation error. There is typically a wide range of parameters that works well.

### 5.2 Neural Network Parameters

In each iteration of the batch gradient descent, we use all the data samples in $I_{train}$. We typically perform 10 iterations of all samples. The weights **W** and biases **b** are initialized by sampling from a normal distribution with mean 0 and standard deviation 0.1. The parameter $C_p$ is set to 100 and the learning rate $\alpha$ is set to 0.0001. The learning process is done offline and it can take up to one hour of execution time in MATLAB to perform 10 iterations of batch gradient descent for 1000 data samples. After the weights and biases have been learned, computing the score for a shape is interactive as this only requires straightforward forward propagations.

### 5.3 Qualitative Patterns in Results

Figure 1 shows the results for a large set of club chairs. We can observe some clear patterns in these results. The highest ranked chair models tend to have more curved surfaces and/or tall (but not too tall) backs. The lowest ranked models tend to have more planar surfaces and the lowest few in particular are somewhat ugly. All of these aspects are learned autonomously from the human aesthetics data and no geometric features that for example correspond to curved, round, or planar surfaces are specified.

There are a few examples of models in the club chair dataset that are the same and they are ranked beside each other. For the fourth chair in the first row of Figure 1, for example, there are a few other chairs that are similar but slightly different. While they are not ranked immediately beside each other, all of them are still ranked near the top. For all the classes of models in general, models that are very similar tend to be ranked near each other. A small change to a 3D shape tend to result in a small change to its ranking and this shows that our algorithm is robust.

Figure 5 shows the aesthetics rankings for four classes of shapes. We describe the patterns that we can observe in the images. For the pedestal tables, the top few models have fancy and/or rounded table legs. The middle row has four similar rounded tables near each other. In the last row, there are a few taller tables followed by the most ugly tables at the end. For the mugs, the top models tend to be tall (but not too tall) and not wide. The handle shapes typically match with the corresponding body shapes of the mugs: they are not too thin and neither too big nor too small. Many mugs in the middle are similar, with subtle differences in their shapes and handles in contrast to the top ones. The bottom ones tend to be the opposite: taller, shorter, wider, and/or with a handle that is thick or thin and too large or more rectangular. The last five are relatively ugly and the upside down one is ranked low (there just happened to be an upside down mug in the downloaded dataset). For the lamps, the top ones tend to be rounded in some way and have some spherical or circular shapes. The bottom ones are wider, taller, planar, and/or not symmetric. The last lamp appears to be broken due to the separated parts of the model. For the dining chairs, the top models tend to have nice proportions, somewhat curved backs, and/or some nice patterns on the backs. The bottom models tend to have taller or simpler backs, and/or planar surfaces.

The results for all classes typically show a distribution where there are some particularly aesthetic shapes, some particularly ugly shapes, and many shapes that are in between. If we take two shapes that are relatively far apart in their computed scores, they typically look quite different in their aesthetics. If we take two shapes that are relatively close in their scores, it can sometimes be difficult to say which one is more aesthetic. There can be cases where given the same two shapes to a group of people, half of them may prefer one shape while the other half may prefer the other shape. We show examples of these cases and analyze them in the evaluation section.

### 5.4 Test Data Sets

In addition to ranking many existing shapes for each class, we can take separate testing sets of 3D models and rank them. Figure 6 shows the rankings of ten such models for four classes and these results show similar patterns as in the rankings above. There are then two ugly shapes for each class that we intentionally made and these have the lowest aesthetics scores. The results here also provide a good way to test that these models can be considered ugly for our control questions in the data collection process.

## 6 Quantitative Evaluation

We perform various types of evaluation to gain a better understanding of our method. Throughout this section, we consider the validation datasets $I_{validation}$ as "correct" or ground truth data and use them to evaluate the accuracy of the learned measure. For each data sample in $I_{validation}$, the learned measure is correct if the collected data says shape A is more aesthetic than shape B and our score of shape A is greater than that of B.

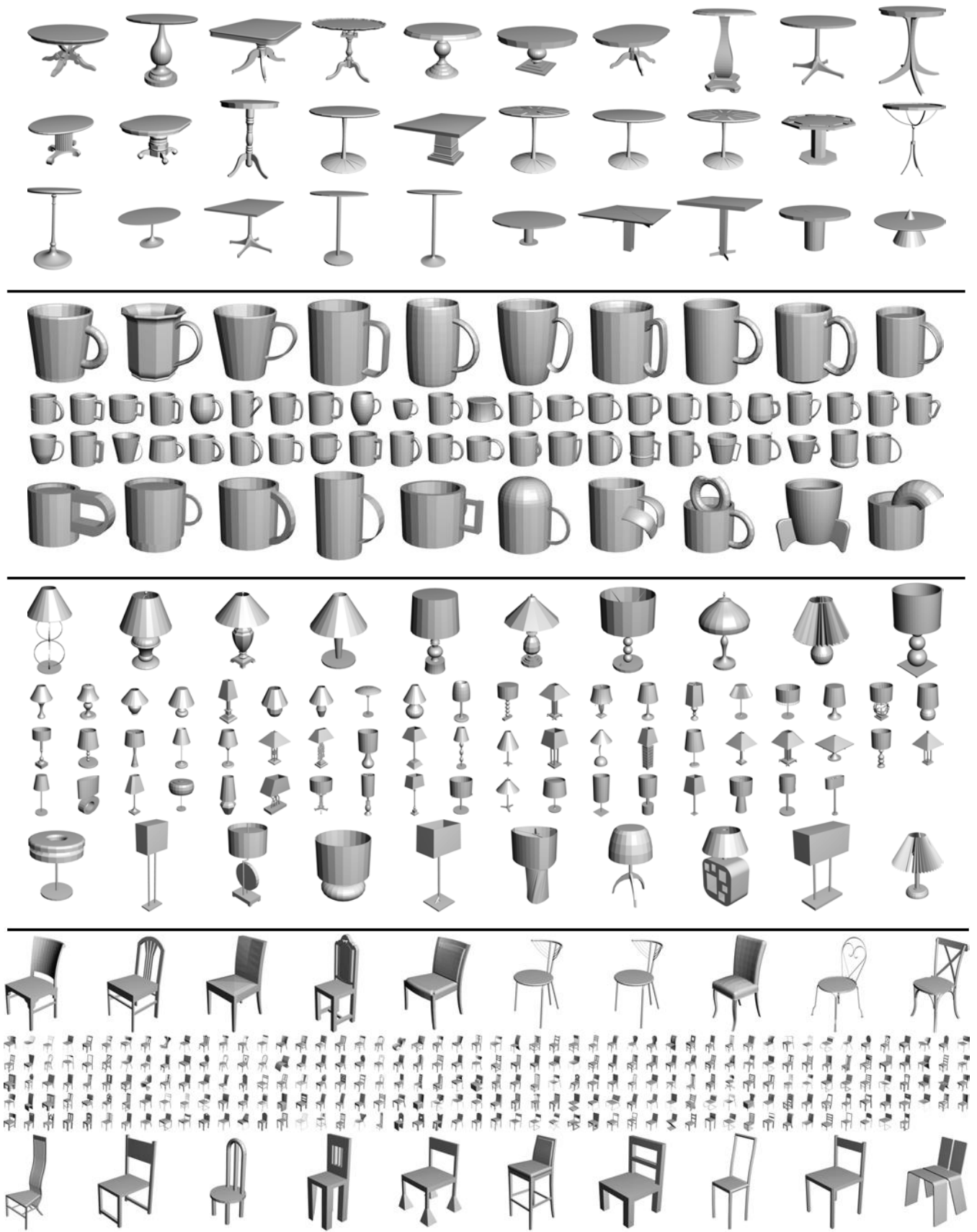

**Figure 5:** *Four sets of 3D shapes ranked (from top to bottom and left to right in each row) according to our aesthetics measure (please zoom in to see shape details). There are 30 pedestal tables, 65 mugs, 78 lamps, and 267 dining chairs.*

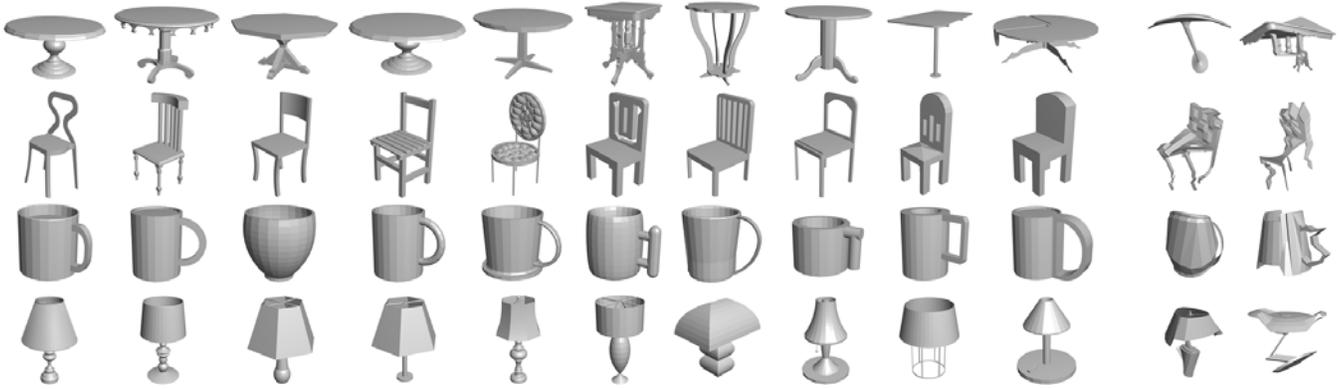

**Figure 6:** *Test sets of shapes ranked (from high to low scores) by our aesthetics measures. There are 4 classes of 10 shapes each: pedestal tables, dining chairs, mugs, and lamps. The last 2 shapes in each row are intentionally created to be ugly shapes and have the lowest scores. The ugly shapes are also used as part of the control questions in the data collection process and are not included in the training data.*

| Class of 3D Shapes | arch full, res 15 | | arch full, res 30 | | arch conv, res 45 | | arch conv, res 60 | |
|---|---|---|---|---|---|---|---|---|
| | all $I_v$ | part $I_v$ | all $I_v$ | part $I_v$ | all $I_v$ | part $I_v$ | all $I_v$ | part $I_v$ |
| Club Chairs | 67.3% | 72.1% | 66.1% | 69.7% | 66.3% | 70.0% | 67.3% | 71.3% |
| Pedestal Tables | 74.7% | 79.1% | 66.3% | 78.5% | 72.1% | 73.5% | 72.4% | 77.2% |
| Mugs | 70.0% | 73.4% | 70.3% | 73.4% | 70.3% | 73.4% | 71.5% | 73.4% |
| Lamps | 64.8% | 74.0% | 61.4% | 72.1% | 66.6% | 74.9% | 66.6% | 73.4% |
| Dining Chairs | 62.4% | 68.3% | 60.9% | 63.9% | 60.4% | 68.1% | 61.6% | 67.4% |

**Table 2:** *Comparison of Network Architectures and Voxel Resolutions. "**arch full**" and "**arch conv**" are the fully-connected and convolutional architectures in Figure 4(a) and (b) respectively. "**res**" is the voxel resolution. "all $I_v$" is where we take all the data samples in $I_{validation}$. "part $I_v$" is where we do not take those data samples in $I_{validation}$ where the difference in aethetics scores between the two shapes is small (i.e. less than ten percent of the range of values). The percentages are the percent of samples that are correctly predicted.*

### 6.1 Comparison of Network Architectures

We show a comparison of different network architectures and voxel resolutions (Table 2). These architectures are non-linear functions that can already represent complex relations from raw voxel data to an aesthetics score, and hence we do not compare them with linear functions. We start with a 15x15x15 voxel resolution for a fully-connected architecture (Figure 4a). This resolution is relatively low but still gives a reasonable representation of the 3D shapes. For this case, there are 200 nodes in the first and second layers and 50 nodes in the third layer. The next resolution is 30x30x30 with the same architecture. As we increase to a 45x45x45 voxel resolution, the training time becomes long and we use a convolutional network (Figure 4b). For this case, the 3D convolution mask in layer 0 is of size 9 (with a stride of 6) and layer 1 is a cube (of nodes) of size 7. The next resolution is 60x60x60 with a convolutional architecture. The motivation for the "part $I_v$" columns is that there are cases where human labels can give uncertain answers (e.g. half chooses A and half chooses B). It is not useful to consider these to measure accuracy as the "ground truth" itself would be uncertain, and it would lead to an unnecessarily lower accuracy.

In Table 2, the percentage accuracies across the architectures are mostly the same. Even if one architecture is better, it is only better by a small amount of a few percent. There is also no clear consensus on which resolution is best. Our analysis shows that a lower resolution already works well. This matches with human intuition as from our experiences, humans only need to observe the overall shapes for a short time to make a decision on their aesthetics. In these experiments, we tested the voxel resolutions of 15 to 60. Reducing the lowest resolution of 15 by half does not work well, as the voxels do not represent the shapes well. Recent work in deep learning for 3D object recognition performs well with a voxel resolution of 24 [Wu et al. 2015]. A resolution of 60 is more than enough to represent the details of many shapes, especially since there is no color information and just the shape geometry.

In addition, taking "part $I_v$" by removing the data pairs with a small difference in their aesthetics scores always improve the percentage accuracies. As these data pairs have a small difference in their scores, we hypothesize that they are also cases where humans can give either answer (i.e. either A or B is aesthetic). Hence if even humans will not be accurate, it can be confusing to include these data samples to measure the accuracy. This justifies that we can use "part $I_v$" as a good indication of accuracy. We provide evidence in the failure and limitation subsection below that data samples where human can give either answer indeed do have small differences in our aesthetics scores.

### 6.2 Quantity of Training Data

The effectiveness of the learned aesthetics measure depends on the quantity of data. We show the percentage accuracy on $I_{validation}$ as the amount of training data increases for each class of shapes (Figure 7). In each case, we train with the number of data samples for 10 iterations of batch gradient descent and compute the accuracy with the full $I_{validation}$ set. For each class, the voxel resolution (we take the best in each class from the previous subsection even if it is better by only a small amount) and amount of data (four units in the x-axis of the graph) are different. The voxel resolutions for club chairs, pedestal tables, mugs, lamps, and dining chairs are 15, 15, 60, 45, and 15 respectively. Each unit of data is 1900, 640, 180, 550, and 1150 samples respectively. In the graph, the main idea is to show that the plots exhibit decreasing returns. As the amount of data keeps increasing, the percentage increases but this increase will slow down. This general trend is the same for the percent accuracy computed with "part $I_v$" which again is always greater than the accuracy computed with the full $I_{validation}$ set. Observing these plots provides one empirical way of knowing whether we have enough training data.

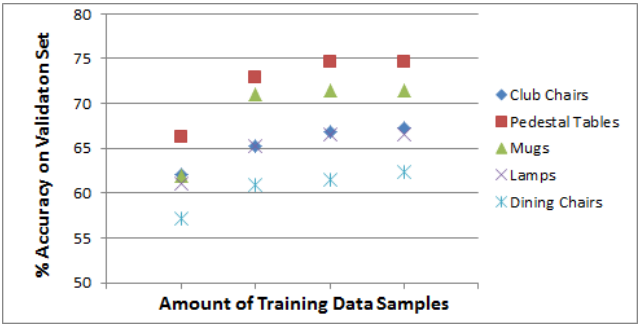

**Figure 7:** *Plots of percent accuracy on $I_{validation}$ versus the amount of data samples in $I_{train}$ for five classes of shapes. For each class, the voxel resolution and amount of data are different, but we plot these on the same graph to show their general trends.*

### 6.3 Analysis of Specific Features

We analyze the learned measure to gain more understanding of what has been learned. There are specific features in 3D shapes that are considered to be important towards aesthetics. For example, shapes that are symmetric and contain many curved surfaces are such features. We show that we have learned these features automatically.

For the symmetry feature, we manually separated 78 lamp models into groups of 67 symmetric and 11 non-symmetric shapes. We consider each shape's rotational symmetry along the up-axis. The mean score for the symmetric group is 0.2031 (Std=0.2286) and for the non-symmetric group is -0.0598 (Std=0.2308). We perform a two-sample t-test assuming equal variances and find a significant effect (t=3.5305; p<0.001), so the population means between the symmetric and non-symmetric groups are different.

For the curvature feature, we manually separated 267 dining chairs into groups of 218 curved and 49 planar shapes. The planar shapes are those that contain only planar surfaces. The mean score for the curved group is 0.0558 (Std=0.1733) and for the planar group is -0.0552 (Std=0.1577). We perform a two-sample t-test assuming equal variances and find a significant effect (t=4.1155; p<0.0001), so the population means between the curved and planar groups are different.

### 6.4 Failure and Limitation Cases

In the data samples, there are some samples where the pairs of shapes have consistent responses. Given the same pair of shapes to different people, they will choose the same response (see examples in Figure 8). In these cases, our aesthetics measure works well. On the other hand, there are some pairs of shapes that can be very close in their aesthetics. Given the same pair of shapes to different people, half of them will choose one shape while half will choose the other (Figure 8). In these cases, our aesthetics measure fails since we will get an accuracy of 50% regardless of what we predict, and a random measure will also get this accuracy. This is the main reason for including "part $I_v$" in our experiments in Table 2, which helps to avoid these cases. This also is the reason that the percentages in Table 2 are reasonable but not high.

We show a numerical analysis of these cases (Figure 9). The idea is to have some samples of pairs of shapes where we collect the aesthetics preferences from multiple people. We then separate the samples into groups based on the multiple responses. For example, a sample pair (A,B) given to 10 people with responses (9,1) (i.e. 9 choose A and 1 choose B), (1,9), or (1,8) goes into the 81-90% group. Note that some Turkers are rejected so we may not have 10 responses for each sample pair. We wish to compute the difference in the aesthetics scores from our learned measure for each sample pair of shapes. If the user chooses A, the difference is the score of A minus the score of B.

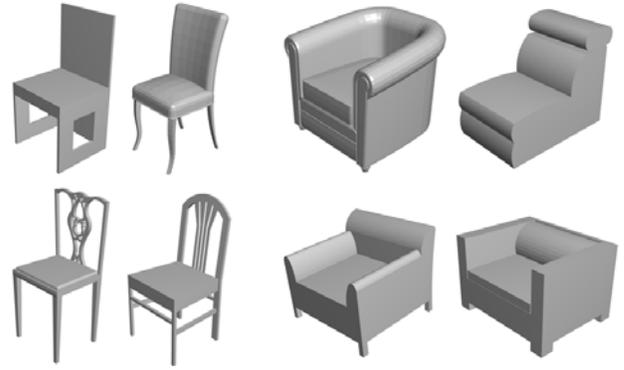

**Figure 8:** *Top: Two example pairs where all ten Turkers chose the same shape (right dining chair and left club chair) as being more aesthetic. Bottom: Two example pairs where five chose one shape and five chose the other.*

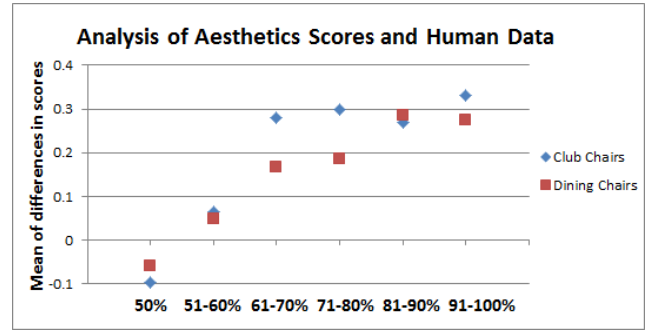

**Figure 9:** *We post 5 HITs and have 10 Turkers provide responses to each HIT. For some HIT tasks, all ten gave the same response (A or B), and these are placed into the 91-100% group. There are some tasks where five chose A and five chose B, and these are placed into the 50% group. For each data sample, we use our learned measure to compute the difference in aesthetics scores. If A is the more common response, we take the score of shape A minus that of shape B. We plot the mean of these differences for each group.*

Figure 9 shows the results for club chairs and dining chairs. We observe an increasing trend in the mean of differences of scores and this trend matches with our intuition. For the 50% group, the two shapes in each pair tend to be similar in aesthetics and the difference in their scores tend to be smaller. If these types of pairs are in $I_{validation}$, it may not be useful to consider them. For the 91-100% group, the two shapes in each pair tend to have a clear difference in aesthetics and the difference in their scores tend to be larger. These are also cases that the learned measure can predict well.

## 7 Applications

The learned aesthetics measure can be used for various applications. We demonstrate the applications of aesthetics-based visualization, search, and scene composition.

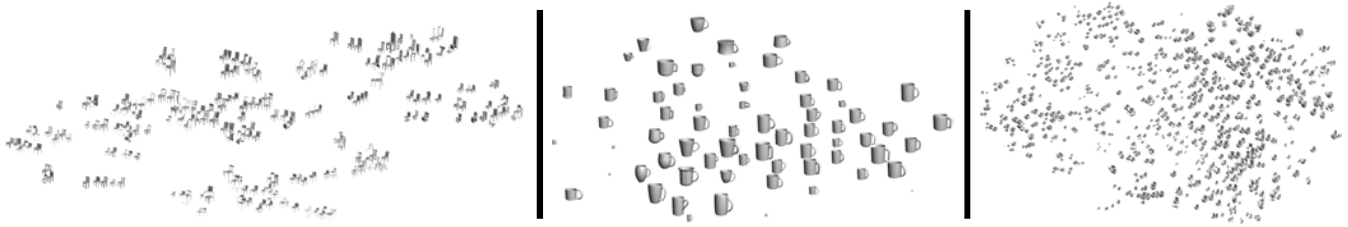

**Figure 10:** *Aesthetics-based Visualization (please zoom in to see shape details). Our aesthetics measure can influence the size of each shape icon and its 2D position within the overall image. We can observe some patterns: for example, there can be regions of shapes similar in both geometry and aesthetics.*

### 7.1 Aesthetics-based Visualization

The idea is to visualize a large dataset of 3D shapes in one image based on aesthetics. First, the aesthetics scores can affect the size of each shape icon in the overall image. We take the aesthetics scores to scale the size of the shape icons. This helps to create a more aesthetics overall image (as the aesthetics shapes are larger) and makes it easier to view the most aesthetic shapes.

Second, the aesthetics measure can affect the 2D positions of the shape icons within the overall image. We can take a voxel resolution of 15 and use t-SNE [van der Maaten and Hinton 2008] to map the raw voxel data to two dimensions. Since this is based on the raw voxels, aesthetics is not considered in this case. If we increase the voxel resolution to 30, t-SNE typically does not work well and the shape icons become mostly laid out with uniform spacing in the overall image. In this case, we take the activation values in the neurons of an inner layer of the neural network (which can be considered as a dimension reduction based on aesthetics) and use t-SNE to map these values to two dimensions. This works well and in this case the aesthetics features learned in the inner layers can influence the 2D positions.

Figure 10 shows some examples of visualizations created this way. We can observe some patterns in these images. For dining chairs, a voxel resolution of 15 works well as input to t-SNE. There are many regions of similar shapes grouped together. Chairs with taller backs tend to be near the top, and there is a group of taller back and aesthetic chairs near the top middle of the overall image. For mugs, taking a voxel resolution of 15 and the 200 activation values of the first layer of the architecture in Figure 4a works well. There are aesthetic mugs around the middle column part of the image while the ugly mugs are very small. The shorter mugs tend to be near the top and the taller mugs tend to be near the bottom. The mugs on the right side of the image tend to have a body shape that is more vertical or cylindrical. For club chairs, taking a voxel resolution of 60 and the activation values of a later layer (e.g. fourth to sixth layer) of the architecture in Figure 4b works well. The more aesthetic club chairs tend to be on the right side of the image. Some chairs with tall backs and curved surfaces are near the bottom right while some chairs with curved arms are near the top right of the image.

### 7.2 Aesthetics-based Search and Scene Composition

We built a search tool where we can rank each class of shapes according to the aesthetics scores. The idea is that it would be easier to browse through and choose from the most aesthetic shapes that are at the top. The ranking results are already shown in the images in the results section. Figure 11 shows some example screenshots of our search tool.

We can use our search tool for 3D scene composition. The idea is that the tool makes it easier to compose more aesthetic shapes together if there are a large number of shapes in each class. Figure 11 shows some example scenes made with our tool.

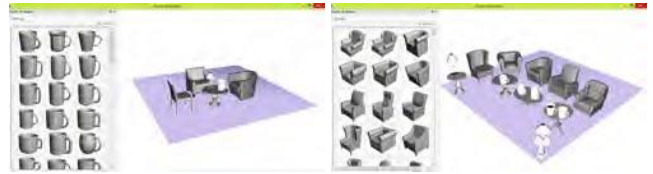

**Figure 11:** *Aesthetics-based Search and Scene Composition. Our search tool displays each class of 3D shapes in the left panel and they can be ranked according to our aesthetics scores. We can use the tool to compose 3D scenes (two examples in image) with the most aesthetic shapes near the top of the left panel. The user can choose how to combine them into a scene.*

## 8 Discussion

We have explored the problem of 3D shape aesthetics and have shown that we can take large datasets of 3D shapes and learn an aesthetics measure from raw voxel data without pre-specifying manually-crafted shape descriptors. The learned "measure" is not a formal measurement but is based on human perception as the data is collected based on human perception. The learned measure is based on data from many people and there can be cases where one person's perception may not agree with it.

To train our aesthetics measures, we take shapes from human-understandable classes and only compare shapes within the same class. However, we still have the "apples and oranges" problem to some extent. For example, our dataset of club chairs is large and there are many variations of club chairs, so there can be different sub-classes of club chairs (or any class).

The attractiveness of an object can be influenced not only by its shape but also by other attributes. These attributes include the color, texture, lighting, and material of 3D shapes [Jain et al. 2012].